\documentclass[a4paper,conference]{IEEEtran}

\usepackage{cite}      

\usepackage{graphicx}  

\usepackage{subfigure} 
\usepackage{aas_macros}
\usepackage{verbatim}
\usepackage{amsmath}
\usepackage{amssymb}
\usepackage{bm}

\newcommand{\etaD}{$\eta_{v\times r}$}
\newcommand{\etaDs}{$\eta_{v\times r}$ }
\newcommand{\etaDm}{\eta_{v\times r}}
\newcommand{\etaP}{$\eta_a\!+\!\eta_b$}
\newcommand{\etaPs}{$\eta_a\!+\!\eta_b$ }
\newcommand{\etaPm}{\eta_a\!+\!\eta_b}
\newcommand{\etaS}{$\eta_{ab}$}
\newcommand{\etaSs}{$\eta_{ab}$ }

\newcommand{\arPs}{Price et. al. (2017) }

\newcommand{\arTPs}{Tricco \& Price (2013) }

\usepackage{fancyheadings}
\begin{document}

\title{Investigating prescriptions for artificial resistivity in smoothed particle magnetohydrodynamics}

\author{\IEEEauthorblockN{James Wurster \& Matthew R. Bate}
\IEEEauthorblockA{School of Physics\\
University of Exeter\\
Exeter, EX4 4QL, United Kingdom\\
j.wurster@exeter.ac.uk}
\and
\IEEEauthorblockN{Daniel J. Price}
\IEEEauthorblockA{Monash Centre for Astrophysics \\
School of Physics \& Astronomy\\
Monash University\\
Clayton, Vic, 3800,  Australia}
\and
\IEEEauthorblockN{Terrence S. Tricco}
\IEEEauthorblockA{Canadian Institute for Theoretical Astrophysics\\
University of Toronto\\
Toronto, Ontario, M5S 1A1, Canada}}


\maketitle
\begin{abstract}

In numerical simulations, artificial terms are applied to the evolution equations for stability.  To prove their validity, these terms are thoroughly tested in test problems where the results are well known.  However, they are seldom tested in production-quality simulations at high resolution where they interact with a plethora of physical and numerical algorithms.  We test three artificial resistivities in both the Orszag-Tang vortex and in a star formation simulation.  From the Orszag-Tang vortex, the \arPs artificial resistivity is the least dissipative thus captures the density and magnetic features; in the star formation algorithm, each artificial resistivity algorithm interacts differently with the sink particle to produce various results, including gas bubbles, dense discs, and migrating sink particles.  The star formation simulations suggest that it is important to rely upon physical resistivity rather than artificial resistivity for convergence.
\end{abstract}

\section{Introduction}
\label{sec:intro}
When applying artificial dissipation terms to the smoothed particle magnetohydrodynamics (SPMHD) equations, one must be careful to apply enough such that the simulation is stable, but not too much such that the results become dominated by the artificial terms.  Moreover, the artificial corrections should only be applied in regions where it is required (e.g. at shocks), and at the minimal amount required for accurate capturing of shocks and other discontinuities.  Thus, determining how, where and when to apply the artificial dissipation terms can be a difficult undertaking.

Several switches to reduce the dissipation away from smooth flow have been derived and tested within the literature.  This includes first- (e.g. \cite{MorrisMonaghan1997}) and second- (e.g.\cite{CullenDehnen2010}) order algorithms for artificial viscosity to be added to the momentum equation; artificial conductivity (e.g. \cite{WadsleyVeeravalliCouchman2008}, \cite{Price2008}) to be added to the energy equation; and artificial resistivity (e.g. \cite{PriceMonaghan2004},\cite{PriceMonaghan2005}, \cite{TriccoPrice2013}, \cite{Phantom2017all}) to be added to the induction equation for magnetic stability.  

Testing the artificial terms is implicitly or explicitly done in most code/algorithm papers.  This is typically done by showing how well the algorithm performs on well-defined tests (e.g. \cite{Sod1978}, \cite{BrioWu1988},  \cite{RyuJones1995}), or on simple problems (e.g. \cite{OrszagTang1979}, \cite{BalsaraSpicer1999}).  During development, these tests are useful for debugging purposes.  After development, the tests are useful for benchmarking purposes, to show proof-of-concept that the algorithm works, to (cautiously) compare the algorithm to other algorithms, and to determine the limitations of the algorithm (c.f. \cite{StoneNorman1992}).

As expected, most codes perform poorly on test problems involving discontinuities when no artificial dissipation terms are applied; with the correct amount of artificial dissipation in the correct place, the numerical results can be made to agree well with the analytical answers (e.g. shock tubes).  It should be noted that when shock tube tests are presented in the literature, they are often performed with maximal artificial corrections rather than the default values.  This difference is demonstrated in figures of 29 and 30 of \cite{Phantom2017all} which demonstrates the Brio \& Wu shock tube \cite{BrioWu1988} using both maximal and default artificial corrections.  By contrast, tests involving smooth flows yield better results when no artificial dissipation is applied (e.g. the advection of a current loop \cite{Price2012}).

Although these artificial dissipations are rigorously tested in test problems, they are seldom tested in realistic or production-quality simulations due to their expense.  However, some artefacts of the artificial algorithms may only appear at high resolution, or once all the required physics is included.  Thus, coupled with the results from test problems, comparing artificial terms in production-quality simulations is required to fully understand the effects of the terms.

In this proceeding, we first discuss smoothed particle magnetohydrodynamics with the focus on magnetic fields and artificial resistivity (c.f. Section~\ref{sec:methods}).  In Section~\ref{sec:test:OT}, we test the three resistivities using the Orszag-Tang vortex test problem and in Section~\ref{sec:test:SF} we test the resistivities in a realistic star formation simulation.  In Section~\ref{sec:optimal} we discuss possible modifications to the resistivities, and we conclude in Section~\ref{sec:conc}.

\section{Numerical Methods}
\label{sec:methods}
\subsection{Smoothed particle magnetohydrodynamics}
The continuum and numerical equations describing smoothed particle magnetohydrodynamics are readily found in the literature (e.g. \cite{Price2012},\cite{WPB2016},\cite{LewisBateTricco2016},\cite{Phantom2017all}).  To evolve the magnetic field, the stress tensor, $S$, used to update the velocity is augmented from $S^{ij} = -P\delta^{ij}$, where $P$ is the gas pressure and $\delta^{ij}$ is the Kronecker delta, to
\begin{equation}
\label{eq:stress}
S^{ij} = -\left(P + \frac{B^2}{2\mu_0}\right)\delta^{ij} + \frac{B^iB^j}{\mu_0},
\end{equation}
where $B$ is the magnetic field and $\mu_0$ is the permeability of free space. Here, $i$ and $j$ represent summation over dimensions $\{x,y,z\}$.

The evolution of the magnetic field is given by the induction equation.  In the continuum limit, this is given by
\begin{equation}
\label{eq:induction}
\frac{{\rm d} B^i}{\text{d} t} = \left(B^j \nabla^j\right)v^i-B^i\left(\nabla^j v^j\right),  \\
\end{equation}
where $v$ is velocity.  The discretised form for the evolution of the magnetic field on particle $a$ is
\begin{eqnarray}
\frac{\text{d} B^i_a}{\text{d} t}  = -\frac{1}{\Omega_a \rho_a} \sum_b m_b & \left[  v^i_{ab} B^j_a \nabla^j_a W_{ab}\left(h_a\right) \right. & \notag \\
&-  \left. B^i_a v^j_{ab} \nabla^j_a W_{ab}\left(h_a\right) \right], &
\end{eqnarray}
where we sum over all particles $b$ within the kernel radius, $W_{ab}$ is the smoothing kernel,  $\rho_a$ is the density, $v^i_{ab} \equiv v^i_a - v^i_b$, and $\Omega_a$ is a dimensionless correction term to account for a spatially variable smoothing length $h_a$ (\!\!\cite{Monaghan2002},\cite{SpringelHernquist2002}).

Magnetic fields have the physical constraint that monopoles do not exist, i.e. $\nabla^i B^i = 0$.  However, this constraint is not explicitly enforced in SPMHD, thus $\nabla^i B^i \ne 0$ can be numerically obtained which can trigger the tensile instability when $\frac{1}{2}B^2 > P$.  The simplest method to correct for this is to subtract $\nabla^i B^i$ from the momentum equation viz.
\begin{eqnarray}
\label{eq:subdivB}
\frac{\text{d} v^i_a}{\text{d} t} \rightarrow \frac{\text{d} v^i_a}{\text{d} t} - f_a B^i_a \sum_b m_b &\left[\frac{B^j_a}{\Omega_a \rho_a^2} \nabla^j_a W_{ab}(h_a) \right. & \notag\\ 
+ &\left.\frac{B^j_b}{\Omega_b \rho_b^2}\nabla^j_a W_{ab}(h_b)\right], &
\end{eqnarray}
using $f_a=1$.  Since this subtraction violates energy and momentum conservation (but only insofar as the divergence term the momentum equation is non-zero; e.g. \cite{Price2012}, \cite{TriccoPrice2012}), it must be treated with caution.  A variable $f_a \in [0,\frac{1}{2}]$ has been suggested \cite{BorveOmangTrulsen2004}, however it has been shown that numerical artefacts can be produced for $f < 1$, thus a more conservative suggestion is $f_a=1$ everywhere \cite{TriccoPrice2012}.  Since the tensile instability is only triggered for $\frac{1}{2}B^2 > P$, we use
\begin{equation}
\label{eq:fdivB}
f_a = \left\{ \begin{array}{l l} 1 ; 	           &  \beta_a \le 1 , \\
                                         2 - \beta_a;  &  1 < \beta_a \le 2, \\
                                         0;                 &  \beta_a > 2,
\end{array}\right.
\end{equation}
where $\beta _a = \frac{2P_a}{B_a^2}$ is the plasma beta.

Finally, artificial resistivity is required to correctly capture shocks and discontinuities.  Thus, the magnetic field evolution is augmented to 
\begin{eqnarray}
\label{eq:B}
\frac{\text{d}B^i}{\text{d}t} &=& \left.\frac{\text{d}B^i}{\text{d}t}\right|_\text{ideal}  + \left.\frac{\text{d}B^i}{\text{d}t}\right|_\text{art},
\end{eqnarray}
where the first term on the right-hand side is the ideal magnetohydrodynamic (MHD) term given in \eqref{eq:induction} and the second term is the artificial resistivity, which is typically represented in the form $\bm{\nabla}\times \left[\eta \left(\bm{\nabla} \times \bm{B}\right)\right]$.  Three possible resistivities are described below, and all three are second-order accurate away from shocks.

\subsection{\arPs artificial resistivity: \etaD} 

This artificial resistivity is the default in \textsc{Phantom} \cite{Phantom2017all} and was used in the study to investigate binary star formation \cite{WPB2017}.  The discretised form of the artificial resistivity (\!\!\cite{PriceMonaghan2004}, \cite{PriceMonaghan2005}) is
\begin{eqnarray}
\left.\frac{\text{d} B^i_a}{\text{d} t}\right|_\text{art}  = \frac{\rho_{a}}{2} \sum_{b} m_{b} \alpha^\text{B} v_{\text{sig},ab}^\text{P}B^i_{ab} & \left[  \frac{ \hat{r}^j_{ab} \nabla^j_aW_{ab}(h_{a})}{\Omega_{a}\rho_{a}^{2}} \right.  &\notag \\
& \left. +  \frac{ \hat{r}^j_{ab}\nabla^j_aW_{ab}(h_{b})}{\Omega_{b}\rho_{b}^{2}} \right], &\label{eq:artificialB}
\end{eqnarray}
where  $\alpha^\text{B} \equiv 1$ is a dimensionless coefficient constant for all particles, $B^i_{ab} \equiv B^i_a - B^i_b$, and the signal velocity is  $v_{\text{sig},ab}^\text{P} = |\bm{v}_{ab} \times \hat{\bm{r}}_{ab} |$.

The main point is that this artificial resistivity is second-order accurate away from shocks since the coefficient is $\eta^\text{P}_{a} = \alpha^\text{B}| \bm{v}_{ab} \times \hat{\bm{r}}_{ab} | h_a \propto h_a^2$.  This algorithm was tested in \cite{Phantom2017all} and shown to still provide sufficient dissipation at magnetic discontinuities.

\subsection{\arTPs term-averaged artificial resistivity: \etaP}
\label{sec:etaP}
This version of artificial resistivity was included in a previous (private) version of \textsc{Phantom} and was used in the study to investigate isolated star formation \cite{WPB2016}.  The discretised form of the artificial resistivity is
\begin{eqnarray}
\left.\frac{\text{d} B^i_a}{\text{d} t}\right|_\text{art}  = \frac{\rho_{a}}{2} \sum_{b} m_{b} B^i_{ab} & \left[  \frac{ \alpha^\text{B}_a v_{\text{sig},a}^\text{T} \hat{r}^j_{ab} \nabla^j_aW_{ab}(h_{a})}{\Omega_{a}\rho_{a}^{2}} \right.  &\notag \\
& \left. +  \frac{ \alpha^\text{B}_b v_{\text{sig},b}^\text{T} \hat{r}^j_{ab}\nabla^j_aW_{ab}(h_{b})}{\Omega_{b}\rho_{b}^{2}} \right], &\label{eq:artificialB}
\end{eqnarray}
where the signal velocity is $v_{\text{sig},a}^\text{T} = \sqrt{c_{\text{s},a}^{2} + v_{\text{A},a}^{2}}$ where $c_{\text{s},a}$ is the sound speed and $v_{\text{A},a}$ is the Alfv{\'e}n velocity, and  $\alpha_a^\text{B} = \min\left(h_a \left|\nabla \bm{B}_a\right|/\left|\bm{B}_a\right|,1\right)$.  This ensures that resistivity is only strong where there are strong gradients in the magnetic field \cite{TriccoPrice2013}. 

The resistivity is also second-order away from shocks, since $\eta^\text{T}_a \approx \frac{1}{2} \alpha^\text{B}_a v^\text{T}_{\text{sig},a} h_a$ and $\alpha^\text{B}_a \propto h_a$.  Note that this coefficient can be calculated without any knowledge of the $a$'s neighbours.

\subsection{\arTPs variable-averaged resistivity: \etaS}
The final artificial resistivity we test is similar to that used in \textsc{sphNG} (\!\!\cite{Benz1990}, \cite{BenzEtAl1990}, \cite{BateBonnellPrice1995}) which was used in the study to investigate isolated star formation (\!\!\cite{LewisBatePrice2015},\cite{LewisBate2017}).  This algorithm has been incorporated into \textsc{Phantom} for this study. This version uses the same signal velocity, dimensionless coefficient and resistivity coefficient as in Section~\ref{sec:etaP} above, but averages the terms differently; various averaging algorithms for the smoothing length have previously been studied \cite{LewisBatePrice2015} but found to be irrelevant.  This artificial resistivity is given by
\begin{eqnarray}
\left.\frac{\text{d} B^i_a}{\text{d} t}\right|_\text{art}  &= \frac{\rho_{a}}{2} \sum_{b} m_{b} B^i_{ab}  \left[\frac{\left(\alpha^\text{B}_a+\alpha^\text{B}_b\right)\left(v_{\text{sig},a}^\text{T} + v_{\text{sig},b}^\text{T}\right)}{\left(\rho_a + \rho_b\right)^2}\right] & \notag \\
& \times \frac{1}{2}\left[\frac{\hat{r}^j_{ab} \nabla^j_aW_{ab}(h_{a})}{\Omega_{a}} +\frac{\hat{r}^j_{ab} \nabla^j_bW_{ab}(h_{b})}{\Omega_{b}}\right].  &
\end{eqnarray}

\section{Idealised test: Orszag-Tang vortex}
\label{sec:test:OT}
Standard tests for MHD include shock tubes (e.g. \cite{BrioWu1988}, \cite{RyuJones1995}), the Orszag-Tang vortex \cite{OrszagTang1979}, and the MHD rotor test \cite{BalsaraSpicer1999}.  We perform the Orszag-Tang test using the SPMHD code \textsc{Phantom} \cite{Phantom2017all} with an adiabatic equation of state with $\gamma = 5/3$.

The particles are placed on a periodic, close-packed lattice with $x, y \in [-0.5,0.5]$ and $z \in [-\frac{\sqrt{6}}{256},\frac{\sqrt{6}}{256}]$; our resolution is $512 \times 590 \times 12$ particles.  The particles have an initial velocity and magnetic field of $[v_x,v_y,v_z] = [-v_0 \sin(2\pi y'), v_0 \sin(2\pi x'), 0.01v_0]$ and $[B_x,B_y,B_z] = [-B_0 \sin(2\pi y'),B_0 \sin(4\pi x'),0]$, respectively, where $v_0 =1$, $B_0 =1/ \sqrt{4\pi}$, $x' =x-x_\text{min}$ and $y' =y-y_\text{min}$.  The initial plasma beta and Mach number are $\beta_0 = 10/3$ and $\mathcal{M}_0 = v_0/c_{\text{s},0} = 1$, respectively, which yield an an initial pressure and density of $P_0 = \frac{B_0^2}{2\beta_0} \approx 0.133$ and $\rho_0 = \gamma P_0\mathcal{M}_0 \approx 0.221$, respectively.  Physical units are irrelevant for this test.

\begin{figure}
\centering
\includegraphics[width=\columnwidth]{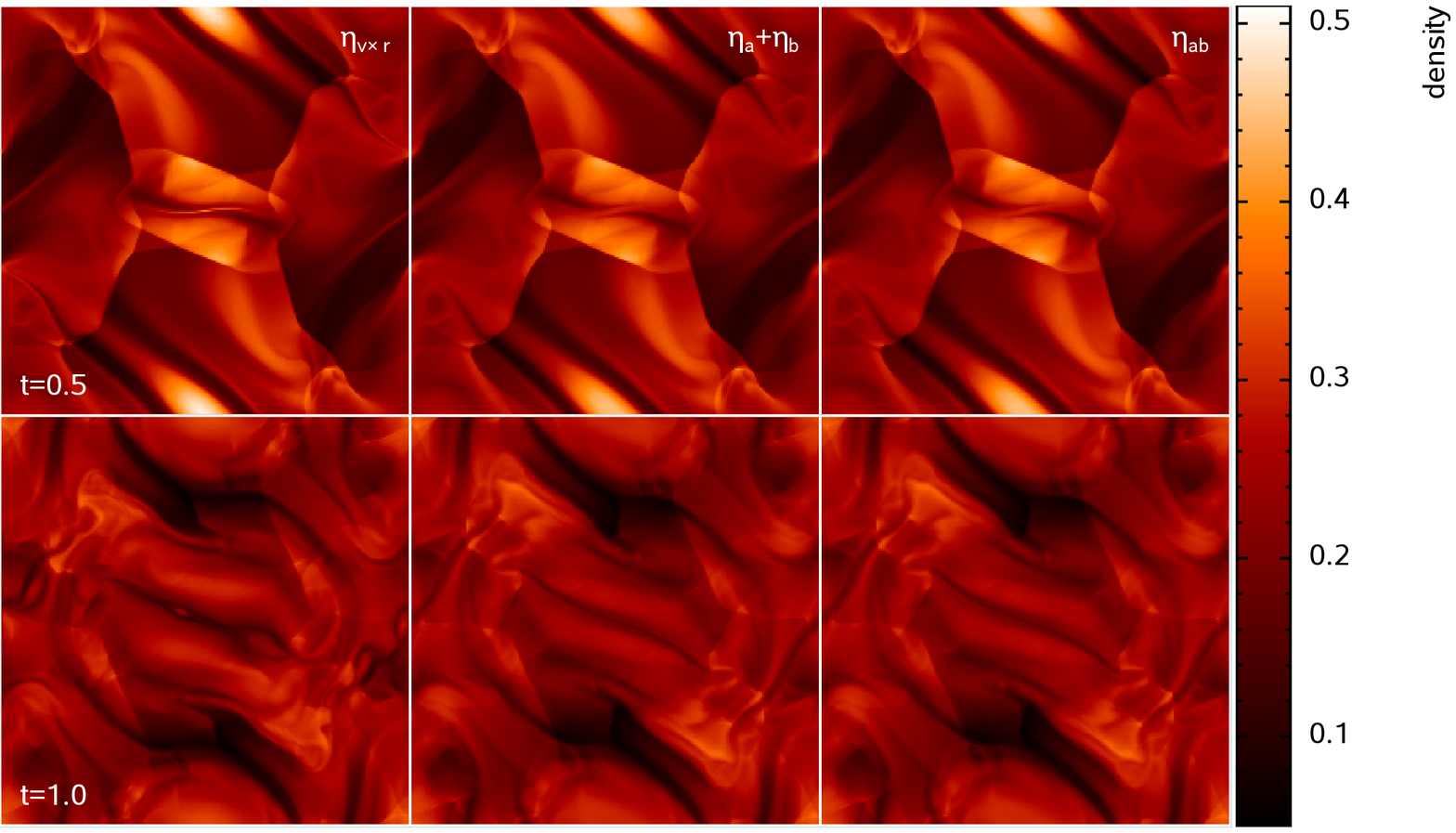}
\caption{The mid-plane gas density for the Orszag-Tang vortex at times $t=0.5$ (top) and $t=1$ (bottom); the initial conditions are given in text.  The left-hand column uses the \arPs artificial resistivity \etaD, the middle column uses the \arTPs term-averaged resistivity \etaP, and the right-hand column uses the \arTPs variable-averaged resistivity \etaS.  The results using \etaPs or \etaSs yield features that are slightly more washed out than using \etaD.  The results from the  \etaPs and \etaSs models are indistinguishable from one another.}
\label{fig:OT:rho}
\end{figure}

Figure~\ref{fig:OT:rho} shows the mid-plane gas density at $t=0.5$ (top) and $t=1$ (bottom) for the three models.
At both times, the features are sharper when using the \etaDs  resistivity; a magnetic island \cite{PolitanoPouquetSulem1989} appears near the centre of the \etaDs model at $t = 1$, but not in the  \etaPs or \etaSs models.  As previously shown in the literature, decreasing the resolution of the \etaDs model removes the magnetic island \cite{Phantom2017all}, thus switching to  \etaPs  has a similar effect to reducing the resolution of the \etaDs model.

\begin{figure}
\centering
\includegraphics[width=\columnwidth]{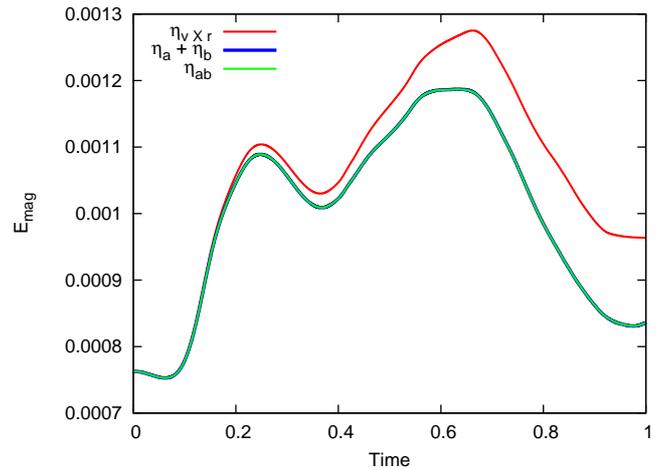}
\caption{The evolution of the total magnetic energy for the  three models presented in code units.  The \etaDs model yields up to 16 per cent less dissipation (at $t\approx0.97$) two the other two models.  The total magnetic energies in \etaPs and \etaSs differ by less than 0.025 per cent at any given time.}
\label{fig:OT:emag}
\end{figure}

Figure~\ref{fig:OT:emag} shows the evolution of the total magnetic energy. 
By design, the \etaPs and \etaSs artificial resistivities are more dissipative, and yield a lower total magnetic energy than \etaD; at $t=0.5$ and 1, the magnetic energy is $\sim4$ and 15 per cent lower, respectively.  At all times, the magnetic energies of \etaPs and \etaSs differ by less than 0.025 per cent; this is expected since there are no steep density gradients.  If the magnetic energies were normalised to their initial value, then the evolution of the total magnetic energy using \etaPs can be approximately matched by using \etaDs  at a resolution of $164 \times 190 \times 12$.

These results suggest that the \etaDs  resistivity yields better results due to its reduced dissipation, and the ability to capture magnetic and density features without needlessly increasing the resolution.

Importantly, models without any artificial resistivity are noticeably worse \cite{Phantom2017all}, suggesting at least a small amount of artificial resistivity is require to capture MHD discontinuities.

\section{Realistic tests: Star formation}
\label{sec:test:SF}
\begin{figure*}[h]
\centering
\includegraphics[width=0.40\textwidth]{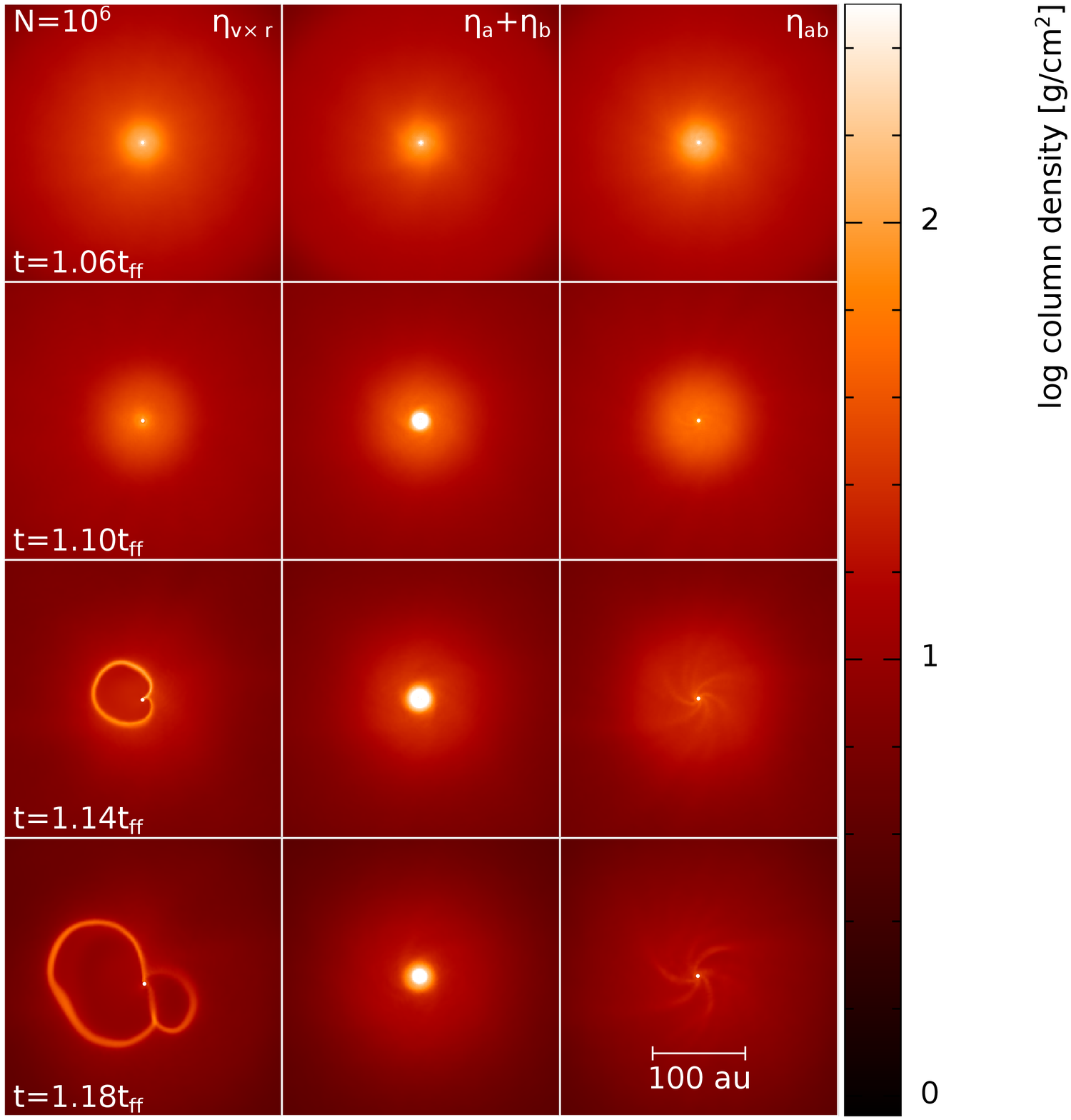}
\includegraphics[width=0.40\textwidth]{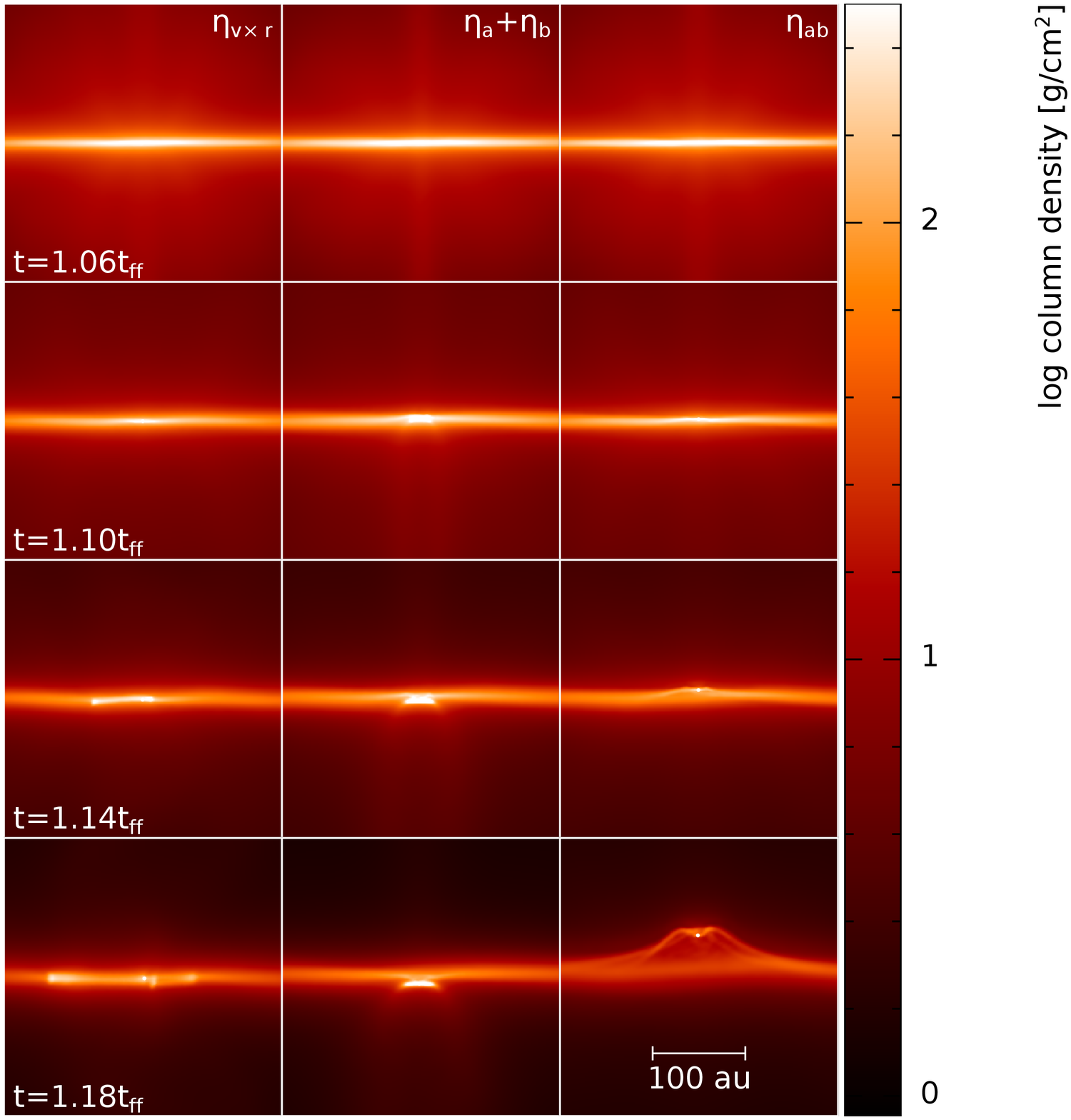}
\caption{Face-on (left) and edge-on (right) gas column density of the isolated star formation model at selected times (in units of the free-fall time, $t_\text{ff} = 2.4\times~10^4$~yr).  The columns show the different models, and the rows show images taken at different times.  The white circles represent the sink particle with the radius of the circle representing the accretion radius of the sink particle.  Each frame is (300~AU)$^2$. At early times, the results are qualitatively model-independent; at late times, the \etaDs model forms bubbles, the \etaPs model has evolved a dense disc, and the \etaSs model's disc has mostly dissipated and its sink particle has levitated due to lack of momentum conservation.}
\label{fig:SF:colden}
\end{figure*}

For a realistic test, we simulate the formation of an isolated protostar as in e.g. \cite{PriceBate2007}, \cite{HennebelleFromang2008}, \cite{DuffinPudritz2009}, \cite{MellonLi2009}, \cite{CommerconEtAl2010}, \cite{LiKrasnopolskyShang2011}, \cite{DappBasuKunz012}, \cite{TomidaOkuzumiMachida2015}, \cite{TsukamotoEtAl2015}, \cite{LewisBatePrice2015},\cite{WPB2016},\cite{LewisBate2017}.  Strongly magnetised numerical simulations of star formation have historically failed to produce discs around protostars, which is contrary to what is observed.  This is known as the magnetic braking catastrophe (e.g. \cite{AllenShuLi2003}, \cite{MellonLi2008}).  This is a result of ideal MHD (where there is no resistivity) efficiently transporting angular momentum away from the protostar, allowing the collapsing gas to be directly accreted onto the protostar rather than first entering a rotationally supported disc. 

We again use \textsc{Phantom} for all three tests, but this time use a Barotropic equation of state, and include gravity and sink particles.  To initialise the problem, we create a spherical cloud of radius $R=4\times~10^{16}$~cm = 0.013~pc, mass $M=1$~M$_{\odot}$, mean density of $\rho_0=7.43\times~10^{-18}$~g~cm$^{-3}$ and $10^6$ SPH particles; we then place the sphere in a low density box of edge length $l = 4R$ and a density contrast of 30:1.  The cloud has an initial rotational velocity of $\Omega = 1.77\times 10^{-13}$ rad s$^{-1}$, and an initial sound speed of $c_\text{s,0} = 2.19\times 10^4$~cm~s$^{-1}$.  The entire domain is threaded with a uniform magnetic field of $B = 163\mu$G which is anti-aligned with the rotation axis.  The free-fall time is $t_\text{ff}=2.4\times~10^4$~yr, which is the characteristic timescale for this study.

To study the long term evolution of the environment after the protostar is formed, sink particles \cite{BateBonnellPrice1995} are required.  These are inserted when the densest gas particle reaches $\rho > \rho_\text{crit} = 10^{-10}$g cm$^{-3}$ and given conditions at and around the sink particle candidate are met.  The sink particles have an accretion radius of $h_\text{acc} = 2$ au; any particle coming within 1 au of the sink is automatically accreted, while particles coming within 2~au are only accreted if given criteria are met.  Accreted particles have their properties added to the sink and then are removed from the simulation.  As per convention, there are no boundary conditions associated with the sink particles.

\begin{figure}
\centering
\includegraphics[width=0.8\columnwidth]{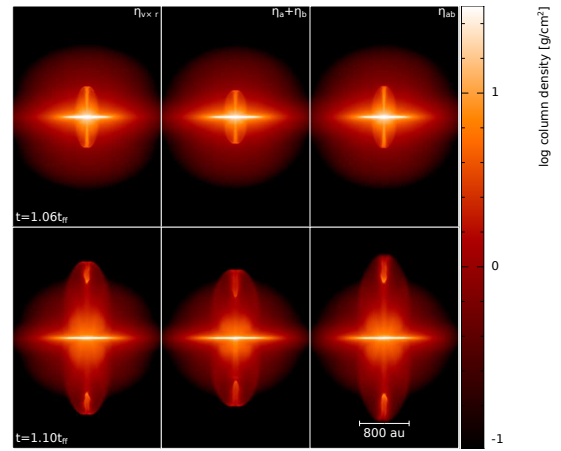}
\caption{Edge-on gas column density at two early times for the star formation simulations, as in Figure~\ref{fig:SF:colden}.  Each frame is $2400\times3600$ au so the large scale structure of the outflows can be seen; the outflows in the \etaPs model are slower then the other two models.}
\label{fig:SF:coldenBig}
\end{figure}

\begin{figure}
\centering
\includegraphics[width=\columnwidth]{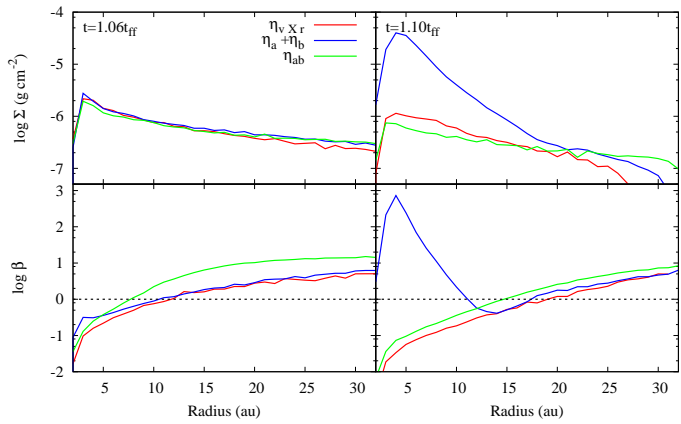}
\caption{The azimuthally averaged gas surface density (top) and plasma beta (bottom) for the inner 32 au at two early times for the star formation simulations.  The horizontal axis starts at $R=2$ au, which is the accretion radius of the sink particle.  If $\beta < 1$, then the gas is dominated by magnetic pressure, otherwise it is dominated by gas pressure.  The results are approximately independent of resistivity model at $t=1.06t_\text{ff}$, however, the \etaPs model diverges from the other two shortly thereafter.}
\label{fig:SF:disc}
\end{figure}

We ran the star formation simulation three times, only changing the artificial resistivity.  Since steep gradients are expected to form near the protostar, it is worth investigating both \etaPs and \etaS, despite them yielding identical results in the Orszag-Tang vortex.  Figure~\ref{fig:SF:colden} shows the face-on and edge-on gas column densities at four different times for the three models, Figure~\ref{fig:SF:coldenBig} shows the edge-on gas column density of the outflows at the first two times, and Figure~\ref{fig:SF:disc} shows the azimuthally averaged gas surface density and plasma beta around the sink particle at the first two times.


At the early time of $t\approx 1.06t_\text{ff}$, the results are similar for all three models.  The surface density profiles differ by at most 40 per cent, and the \etaSs model has a plasma beta that is 2-3 times higher than the \etaDs model.  All but the inner few au are dominated by gas pressure, thus at this time, the magnetic fields are of a secondary importance over most of the domain.  Thus, arguably, if the study were to end here, the specific choice of artificial resistivity would not be important.

As the models continue to evolve, they begin to diverge.  By $t \approx 1.10t_\text{ff}$, there is a dense disc around the sink in the \etaPs model due to its large resistivity, but not the other models; this model also has slower outflow velocities and hence the outflows have not progressed as far as in the other two models.  This is consistent with previous results that showed an inverse correlation between disc size and outflow velocity \cite{WPB2016}.  At this time, the other two models have surface densities and plasma beta's that differ by less than a factor of two; their maximum surface density is $\sim$60 times lower than in the \etaPs model.  At this time, \etaDs  and \etaSs  demonstrate the magnetic braking catastrophe whereas the \etaPs model does not.  Given that this disc is likely formed by the artificial resistivity,  we must be cautious to not claim that the problem has been solved.  At this stage, it is arguable that \etaDs  or \etaSs  are likely preferable algorithms since they yield similar results.

As the \etaPs model continues to evolve, the disc slowly decreases in mass and radius, but the system remains stable.  The disc slightly migrates downwards due to lack of momentum conservation; by $t\approx 1.21t_\text{ff}$, the sink particle has drifted d$z \approx 9.7$ au from its creation point and has a speed of $v_\text{z} \approx 0.03$ km s$^{-1}$.  
Tests have shown that the amount of migration is dependent on sink size, however, low resolution cores without sinks have also been found to migrate.  

As the \etaDs  and \etaSs models evolve, the gas density near the sink decreases, but, as in the \etaPs model, the magnetic field continues to increase.  As a result, the magnetic pressure increasingly dominates gas pressure near the sink in these two models, and stronger magnetic pressure in SPMHD results in lower momentum conservation (i.e. $\beta \ll 1$; see \eqref{eq:subdivB} and \eqref{eq:fdivB}).  This lack of momentum conservation in the \etaSs model results in the sink particle migrating in the vertical direction, which drags the gas with it; azimuthal symmetry is approximately preserved.  Since there is only low-density gas around the sink (as compared to \etaP), the sink particle is capable of wandering with great speeds since there is no gas disc to exert an attractive gravitational force.  By $t\approx 1.21t_\text{ff}$, the sink has a vertical velocity of $v_\text{z} \approx 0.5$ km s$^{-1}$ and moved a distance of d$z \approx 108$ au.

The sink particle in the \etaDs model has a maximum vertical wandering of $v_\text{z} \approx 0.009$ km s$^{-1}$; this is similar to hydrodynamical simulations that do not suffer from an intrinsic lack of momentum conservation.  In this model, the momentum conservation error contributes to the gas motion, rather than the sink particle migration.  Since the dominant velocity component is the radial component, the gas receives a small radial kick.  Coupled with the high magnetic pressure, this ultimately causes the launching of the gas bubbles.

\begin{figure}
\centering
\includegraphics[width=\columnwidth]{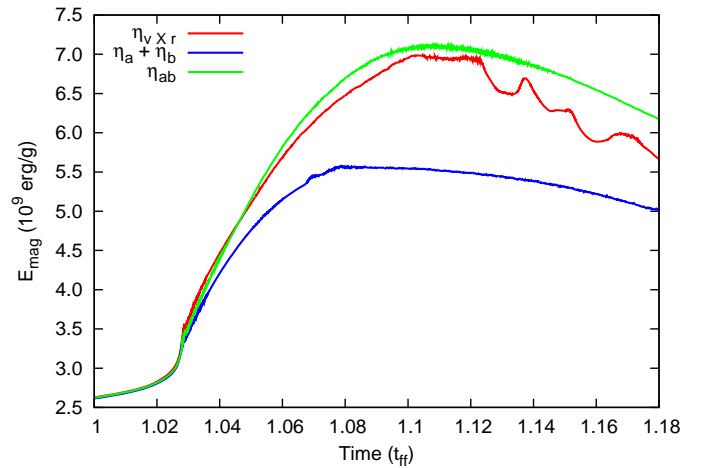}
\caption{The evolution of the total magnetic energy for the star formation simulations.  The three models begin to diverge at $t\approx1.03t_\text{ff}$, which coincides with the insertion of the sink particle.  The \etaPs  is the most resistive, hence the lowest total magnetic energy.  The \etaDs  forms the gas bubbles at $t\approx1.12t_\text{ff}$, when the curve becomes less smooth.}
\label{fig:SF:emag}
\end{figure}

Figure~\ref{fig:SF:emag} shows the evolution of the total magnetic energy.
As with the Orszag-Tang vortex, the \etaPs model shows the most magnetic dissipation.  Thus, although this model is arguably more stable than the other two, it has dissipated enough energy such that the results are at least in part being controlled by the artificial resistivity.

\begin{figure}
\centering
\includegraphics[width=0.40\textwidth]{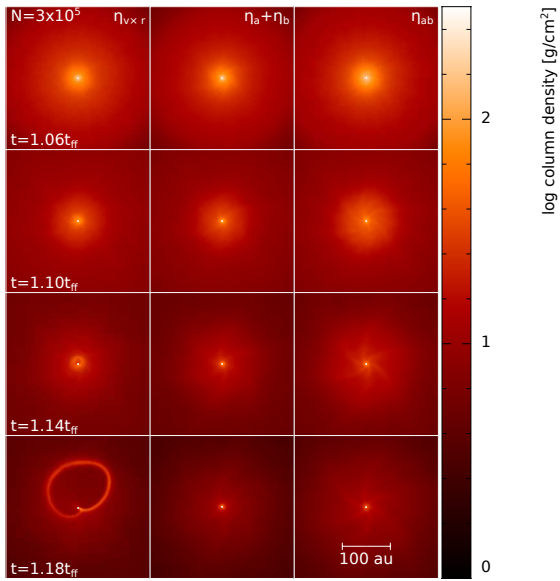}
\caption{Face-on gas column density of the isolated star formation model at selected times as in Figure~\ref{fig:SF:colden} except using $3\times 10^5$ particles in the sphere.  in units of the free-fall time, $t_\text{ff} = 2.4\times~10^4$~yr).  Compared to their higher resolution counterpart, the \etaDs model is delayed in producing the gas bubbles, the \etaPs model has a smaller disc, and the sink in the \etaSs model does not migrate.}
\label{fig:SF:colden:low}
\end{figure}

When we reduce the resolution to $3\times 10^5$ particles in the sphere, there is good agreement amongst the models at early times and their evolution diverges at late times; see Figure~\ref{fig:SF:colden:low}, which shows the face-on gas column density at four times for the low resolution models.
At the lower resolution, the density, magnetic gradients, and the sink boundary are less resolved, thus any features that may cause the evolution to diverge in the higher resolutions models get smoothed out, thus allowing for better convergence.  

As a result, the gas bubbles are launched at a later time in the lower resolution \etaDs model.  The low resolution \etaPs model has the highest surface density amongst the three low resolution runs, but this is $\sim$2 dex lower than in the higher resolution run.  Prior to sink insertion, the higher resolution model has a higher density and central magnetic field simply due to the resolution.  After the sink particle has been inserted, the sink grows more slowly due to the lower gas particle masses and better resolved boundaries.  This leaves more gas in the sink's environment, which ultimately forms a dense disc as the magnetic field is dissipated; the lower resolution model does not have enough gas in the vicinity of the sink to form a dense disc.  In low resolution \etaSs model, the sink migration is minimal.  Finally, since the artificial resistivity is $\eta \propto h^2$, there is more artificial resistivity in the lower resolution models, resulting in lower total magnetic energies.

This study is not the first time that magnetic bubbles have been discovered in MHD simulations.  They have been previously found in similar 3D adaptive mesh refinement (AMR) simulations \cite{KrasnopolskyLiShangZhao2012}, although their bubbles were not confined to the mid-plane.  As with this study, their magnetic bubbles were launched from the low-density gas near the sink particle, and contained strong magnetic fields.  Their study was performed on a spherical-polar grid, thus similar to sink particles in SPH, there is a hole in the magnetic field around the sink particle.  Thus, the same bubbles have been observed using two distinct numerical algorithms, and in both cases they originate in low-density gas near sink particles.

The long term evolution leads to a conundrum of which artificial resistivity to use:  \etaDs  leads to magnetic bubbles launched from near a numerical boundary, but that have been also observed in AMR simulations; \etaPs  leads to the formation of a massive disc; and \etaSs  yields a low mass discs, but ultimately leads to fast migration of the sink particle.  

\section{Determining an artificial resistivity formulation with minimal resistivity and without gas bubbles}
\label{sec:optimal}

To conclusively determine if the gas bubbles are physically or numerically produced, similar simulations without sink particles need to be run for similar lengths of times.  However, this is prohibitively expensive since the central density of the protostar reaches $\rho \sim \mathcal{O}(1)$ g cm$^{-3}$, which is $\sim10$ dex greater than the density in the disc.  Thus, studies that exclude sink particles are evolved for several hundred years after the formation of the protostar (e.g. \cite{BateTriccoPrice2014}, \cite{Tomida2014}, \cite{TsukamotoEtAl2015}), whereas these star formation simulations were evolved for $\sim$5000 years after the formation of the protostar; the gas bubbles formed $\sim$2500 years after the formation of the protostar.  Thus, this problem must be solved using simulations that include sink particles.

Assuming that the gas bubbles are artificial, we ran several star formation simulations with various modifications in attempts to prevent their formation while applying the minimal amount of artificial dissipation possible (i.e. less dissipation than in the \etaPs model).  In order to achieve this, our focus was on increasing the resistivity near the sink particle, but leaving it low elsewhere.  If this could be achieved, then the gas pressure near the sink particle would be comparable or greater than the magnetic pressure (i.e. $\beta \gtrsim 1$), which would lead to better momentum conservation since there would be less numerical $\nabla^iB^i$ to subtract.

\subsection{Position-dependent resistivity algorithm}
For the first attempt, the choice of resistivity algorithm was made to depend on the particle's distance from the sink particle, $\Delta r$.  Specifically, 
\begin{equation}
\label{eq:etaDP}
\eta \rightarrow \left\{ \begin{array}{l l} \etaPm ; 	             & \frac{\Delta r}{h_\text{acc}} < f_1, \\
                                          \frac{f_2 - \frac{\Delta r}{h_\text{acc}}}{f_2-f_1}(\etaPm) +  \frac{\frac{\Delta r}{h_\text{acc}} - f_1}{f_2-f_1}\etaDm; & f_1 <  \frac{\Delta r}{h_\text{acc}} < f_2 ,\\
                                          \etaDm;                    & f_2 < \frac{\Delta r}{h_\text{acc}}, \\
\end{array}\right.
\end{equation}
where $h_\text{acc}$ is the accretion radius of the sink particle and $f_n$ are free parameters.  For $f_1 = 2$ and $f_2=3$, the gas bubbles formed similarly to the \etaDs model since there were not enough particles with $\frac{\Delta r}{h_\text{acc}} r < f_2$ to make a substantial reduction to the magnetic field.  

Increasing $f_n$ must be done with caution; if $f_n$ are too large, then a large fraction of the region of interest may be within modified region, thus susceptible to the numerical artefacts that arise from merging the two resistivity algorithms.  With caution, we tested $f_1 = 5$ and $f_2=10$.  Since $h_\text{acc} = 2$ au, our modified region extended to 20 au; note that the disc in the \etaPs model extended to $r \approx 30$ au.  For any given particle, the two resistivities can differ by a factor of $\sim$100.  Thus, there is a large decrease in dissipation over the transition region.

As the gas collapses inwards it reaches this transition region and stalls since it is now better able to retain its angular momentum.  Evolving from the \etaDs to the \etaPs resistivity causes both a sharp decrease in the magnetic field in the transition region and separates the radial flow into the gas flowing into this stall region and the gas flowing from the stall region onto the sink.   The gas is then slowly accreted onto the sink particle from between the sink and the stall region.  Since the magnetic field does not decrease as the gas density decreases, the magnetic pressure builds up until it causes the gas bubble to form.  

\subsection{Momentum rather than velocity}
In SPH, values are calculated weighted by density.  Thus, we next tried modifying \etaDs so that its signal velocity used momentum rather then velocity, viz.,
\begin{equation}
\label{eq:pxr}
v_{\text{sig},ab} = \frac{|\bm{p}_{ab} \times \hat{\bm{r}}_{ab} |}{\rho_{ab}},
\end{equation}
where $\bm{p}_{ab} = \rho_a\bm{v}_a - \rho_b\bm{v}_b$.  This resistivity also produced the gas bubbles, although their formation was delayed by d$t \approx 0.02t_\text{ff}$ compared to the fiducial \etaDs model.

\subsection{Sink particle size}
Our next attempt was to reduce the size of the sink particle and use \etaDs everywhere.  By reducing its accretion radius, the region around the protostar would be better resolved, which should better capture the behaviour of the magnetic field.  As expected, both density and the magnetic field strength increased near the sink particle, which lead to an increase in magnetic pressure near the sink.  This caused the gas bubbles to form sooner.  

Since the size of the sink particle should not affect the gas far from it (and this has been verified in tests), then this suggests that the gas bubbles are numerical rather than physical in origin.

\section{Conclusion}
\label{sec:conc}
In this proceeding, we have discussed the importance of testing artificial resistivity in both test cases and production-quality simulations.  Using \textsc{Phantom}, we have tested three different resistivities: \arPs artificial resistivity, \etaD; \arTPs term-averaged artificial resistivity, \etaP; and \arTPs variable-averaged artificial resistivity \etaS.  Each of these artificial resistivities have been previously used in the literature.

Our tests of the Orszag-Tang vortex showed that the \etaDs artificial resistivity was the least resistive and could resolve the magnetic islands at the resolution presented; at the same resolution, the total magnetic energies of the \etaPs and \etaSs models were up to 16 per cent higher than the \etaDs model, and these models were unable to resolve the magnetic islands.

Our tests in the star formation simulations showed that the long-term evolution was dependent on the artificial resistivity model.  Gas bubbles were launched from near the sink particle in the model with the \etaDs artificial resistivity; a large, dense disc formed in the model with the \etaPs artificial resistivity; and the sink particle underwent vertical migration with the \etaSs artificial resistivity.  Thus, each of the artificial resistivities yielded conflicting results.  Aside from the lack of convergence, this could also lead the user to reach an incorrect conclusion if only one of the artificial resistivities was used.

Finally, we tried several methods of preventing the gas bubbles from forming while trying to maintain the minimal dissipation of the \etaDs artificial resistivity.  Although some modifications delayed the formation of the gas bubbles, all attempts where the \etaDs artificial resistivity was the dominant artificial resistivity ultimately produced the gas bubbles.

Future attempts to avoid the gas bubbles should include physical resistivity (e.g. \cite{WardleNg1999},\cite{NakanoNishiUmebayashi2002},\cite{TassisMouschovias2007b}, \cite{BraidingWardle2012accretion}), since it is both physically motivated and should be resolution-independent.  Studies have already included physical resistivity into star formation simulations (e.g. \cite{LiShu1996},\cite{Mouschovias1996},\cite{MellonLi2009},\cite{MachidaInutsukaMatsumoto2011},\cite{LiKrasnopolskyShang2011},\cite{DappBasuKunz012},\cite{TomidaEtAl2013},\cite{TsukamotoEtAl2015},\cite{WPB2016}), and none produced gas bubbles.  However, it is unknown if the failure to produce the gas bubbles was a result of the physical resistivity or the choice the artificial resistivity.  Thus, a comparison similar to this proceeding should be carried out where physical resistivity is included.

Thus, although the standard tests are useful for showing how well an artificial resistivity (or algorithm in general) works, artificial resistivity algorithms should also be tested in production-quality simulations to determine their effect when combined with additional physical and numerical algorithms at high resolutions.  The results may be surprising.


\section*{Acknowledgment}
JW and MRB acknowledge support from the European Research Council under the European Community's Seventh Framework Programme (FP7/2007- 2013 grant agreement no. 339248).  We would like to thank Benjamin T. Lewis for useful discussions. The calculations for this paper were performed on the University of Exeter Supercomputer, a DiRAC Facility jointly funded by STFC, the Large Facilities Capital Fund of BIS, and the University of Exeter.  For the column density figures, we used {\sc splash} \cite{Price2007}. 



%

\bibliographystyle{IEEEtran.bst}
\bibliography{IEEEabrv,../../../Papers_Im_Writing//Wurster_bib}

\end{document}